\documentclass[journal]{IEEEtran}
\usepackage{amsmath,amsfonts}
\usepackage{algorithmic}
\usepackage{algorithm}
\usepackage{array}
\usepackage[caption=false,font=normalsize,labelfont=sf,textfont=sf]{subfig}
\usepackage{textcomp}
\usepackage{stfloats}
\usepackage{url}
\usepackage{verbatim}
\usepackage{graphicx}
\usepackage{cancel}
\usepackage{cite}
\begin{document}

\title{Parallel GPT: Harmonizing the Independence and Interdependence of Acoustic and Semantic Information for Zero-Shot Text-to-Speech}

\author{Jingyuan Xing, Zhipeng Li, Jialong Mai, Xiaofen Xing,~\IEEEmembership{Member,~IEEE}, Xiangmin Xu,~\IEEEmembership{Senior Member,~IEEE}
       
\thanks{Jingyuan Xing is with the School of Future Technology, South China University of Technology, Guangzhou 510640, China. (e-mail: 202320163324@mail.scut.edu.cn)}
\thanks{Zhipeng Li, Jialong Mai and Xiaofen Xing are with the School of Electronic and Information, South China University of Technology, Guangzhou 510640, China. (e-mail: 202221013256@mail.scut.edu.cn;202320111090@mail.scut.edu.cn; xfxing@scut.edu.cn)}
\thanks{Xiangmin Xu is with the School of Future Technology, South China University of Technology, Guangzhou 510640, China, and also with Pazhou Lab, Guangzhou 510335, China (e-mail: xmxu@scut.edu.cn)}
\thanks{Corresponding authors: Xiaofen Xing; Xiangmin Xu.}}

\markboth{Journal of \LaTeX\ Class Files,~Vol.~14, No.~8, August~2021}%
{Shell \MakeLowercase{\textit{et al.}}: A Sample Article Using IEEEtran.cls for IEEE Journals}


\maketitle

\begin{abstract}

Advances in speech representation and large language models have enhanced zero-shot text-to-speech (TTS) performance. However, existing zero-shot TTS models face challenges in capturing the complex correlations between acoustic and semantic features, resulting in a lack of expressiveness and similarity. The primary reason lies in the complex relationship between semantic and acoustic features, which manifests independent and interdependent aspects.
This paper introduces a TTS framework that combines both autoregressive (AR) and non-autoregressive (NAR) modules to harmonize the independence and interdependence of acoustic and semantic information. The AR model leverages the proposed Parallel Tokenizer to synthesize the top semantic and acoustic tokens simultaneously. In contrast, considering the interdependence, the Coupled NAR model predicts detailed tokens based on the general AR model's output. Parallel GPT, built on this architecture, is designed to improve zero-shot text-to-speech synthesis through its parallel structure.
Experiments on English and Chinese datasets demonstrate that the proposed model significantly outperforms the quality and efficiency of the synthesis of existing zero-shot TTS models. Speech demos are available at https://t1235-ch.github.io/pgpt/.

\end{abstract}


\section{Introduction}

\IEEEPARstart{T}{ext-to-speech} (TTS) aims to convert text into natural speech. Traditional TTS methods include concatenative synthesis \cite{wei2000corpus} \cite{black97_eurospeech} and statistical parametric synthesis \cite{zen2009statistical} \cite{zen2013statistical} \cite{tokuda2013speech}. With deep learning advancements, neural network-based TTS models \cite{shen2017tacotron} \cite{ren2020fastspeech} \cite{lim22_interspeech} have emerged, generating spectrograms directly from phonemes. Zero-shot synthesis models such as YourTTS \cite{casanova2022yourtts}, StyleTTS \cite{li2022styletts}, and StyleTTS 2 \cite{li2023styletts} have been further advanced, allowing high-quality expressive speech for unseen speakers.

The emergence of self-supervised speech learning (SSL) models, such as Wav2vec \cite{schneider2019wav2vec} and Hubert \cite{hsu2021hubert}, alongside advances in large language models (LLMs) \cite{radford2019language} \cite{touvron2023llama}, has driven a shift in zero shot TTS research toward decomposing the synthesis framework. Unlike traditional sample-by-sample synthesis \cite{van2016wavenet} or direct Mel-spectrogram estimation \cite{shen2017tacotron} \cite{ren2020fastspeech}, recent approaches \cite{zhu2023vec} \cite{kim2023transduce} leverage speech representations to divide the synthesis framework into two stages. The first stage involves predicting intermediate representations from text, while the second stage converts these intermediate representations into speech. In the first stage, LLMs enhance both the expressiveness and flexibility of the synthesis process.

\begin{figure}[!t]
\centering
\includegraphics[width=\columnwidth]{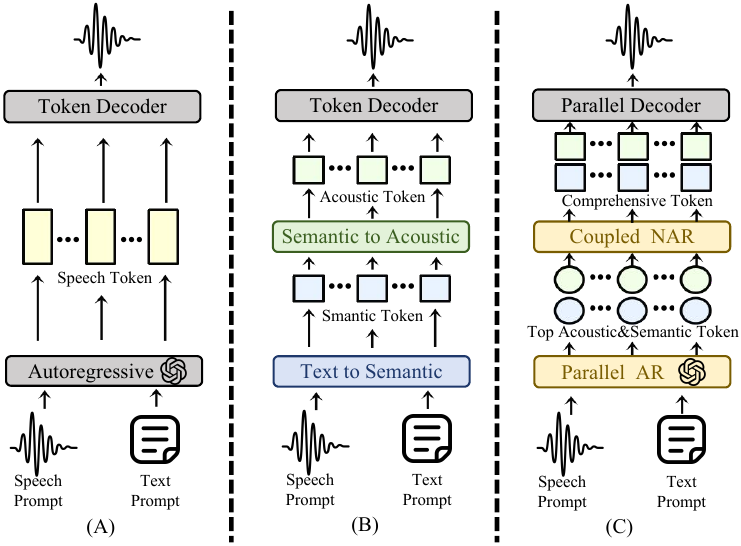}
\caption{Schematic Comparison of Three Speech Synthesis Frameworks. (A) Overall speech token-based synthesis framework, where speech is generated through overall speech tokens. (B) Two-stage semantic-acoustic synthesis framework, where the model generates the semantic tokens first and then synthesizes the acoustic tokens. (C) Parallel GPT-based synthesis framework, where acoustic and semantic tokens are processed simultaneously.}
\label{1}
\end{figure}

Despite advances in zero-shot TTS modeling, several challenges remain. 
Speech encompasses semantic and acoustic information, which are independent in certain aspects but mutually interdependent \cite{fujisaki2004information}. 
LLMs are initially designed for natural language processing (NLP) tasks, emphasizing semantic understanding. However, their direct application to speech synthesis presents challenges in simultaneously understanding both semantic and acoustic information, leading to reduced sensitivity to acoustic features.
It produces unnatural and less diverse speech, particularly when adapting to different speakers or styles.

Challenges in capturing the complex relationship between semantics and acoustics remain prevalent in zero-shot TTS models. These challenges manifest themselves in three key areas: speech synthesis frameworks, speech token construction, and LLM-based methods.

For the frameworks for speech synthesis, traditional models can be categorized into two main types. The first framework directly predicts the entire speech token \cite{wang2023neural} \cite{zhu2023vec}, which is then converted into speech, as shown in Figure \ref{1} (A). 
It is intuitive and directly connects text to speech, but the complexity of speech tokens makes the first-stage prediction challenging, often causing significant information loss. Therefore, researchers have proposed a second framework that first predicts semantic tokens from text and then predicts acoustic tokens from the semantic tokens \cite{wang2024maskgct} \cite{zhang2024speechtokenizer}, as shown in Figure \ref{1} (B). This approach reduces the complexity of the prediction in the first stage.

However, the second method still has limitations. The relationship between acoustics and semantics involves independence and interdependence. 
First, it creates a semantic-driven initial stage. 
Many features with a weaker connection to semantics, such as sentence pauses, breath sounds, sighs, and pitch variation, may be constrained, even though they are closely connected to the acoustics. 
Second, in the subsequent stage,
when semantic tokens are converted into acoustic tokens, some information with a strong semantic connection may be overshadowed by acoustic predictions.
Moreover, the temporal alignment between semantic and acoustic tokens can also impact the efficiency of model predictions.
Consequently, achieving a synthesis framework that integrates semantic and acoustic elements in parallel remains a significant challenge.

For the construction of speech tokens, traditional approaches can be classified into Overall speech tokens, only semantic tokens, and two-stage tokens. The overall speech tokens method \cite{defossez2022high} \cite{zeghidour2021soundstream} \cite{yang2024simplespeech} treats speech as a unified whole, combining all the information in speech. The Only Semantic Tokens method \cite{kim22c_interspeech} \cite{du2024funcodec} \cite{liu2024generative} focuses exclusively on extracting linguistic content, enhancing the clarity and expressiveness of speech.
The Two-Stage Tokens approach \cite{zhang2024speechtokenizer} \cite{wang2024maskgct} separates speech into lower-level semantic tokens and higher-level acoustic tokens, allowing a gradual synthesis of speech by combining these components step by step.

\begin{figure*}[!t]
\centering
\includegraphics[width=\textwidth]{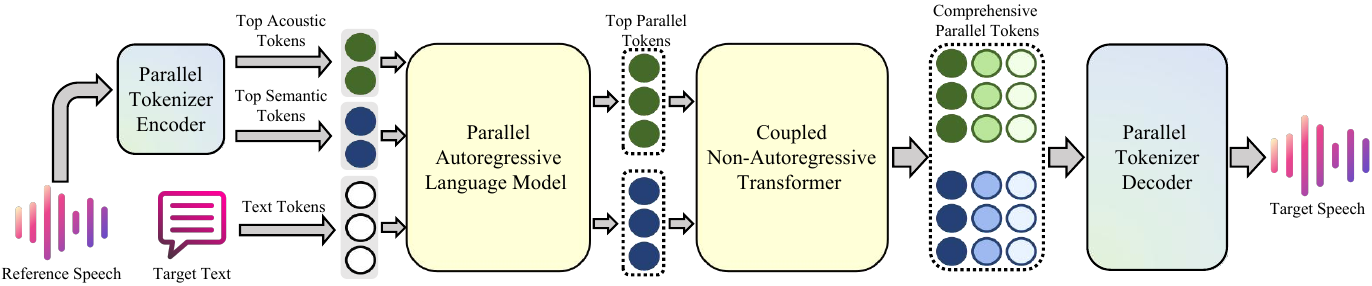}
\caption{An overview of the Parallel GPT inference pipeline. (1) The Parallel Tokenizer Encoder derives tokens from reference speech. (2) The Parallel Autoregressive Language Model generates the top semantic and acoustic tokens based on the condition of text and speech. (3) 
The Coupled Non-Autoregressive Transformer model generates the semantic and acoustic tokens of the last two bits. These outputs, combined with the top parallel tokens, form the comprehensive parallel tokens. (4) The Parallel Tokenizer Decoder generates higher-quality speech from generated comprehensive parallel tokens.}
\label{2}
\end{figure*}

However, these methods still have limitations. They do not fully capture the inherent independence between semantics and acoustics, which can lead to a loss of detail in speech synthesis.
Consequently, it is essential to propose an effective method of constructing speech tokens that harmonizes the independence and interdependence of acoustic and semantic information.

In TTS applications of LLMs,
AR-based models \cite{radford2019language} \cite{touvron2023llama} demonstrate superior diversity, prosody, expressiveness, and flexibility due to implicit duration modeling and autoregressive sampling strategies \cite{shen2018natural}.
In these models \cite{wang2023neural} \cite{yang2023uniaudio} \cite{guo2024socodec}, LLMs predict speech or semantic tokens, enhancing linguistic fluency and context awareness by fine-tuning or retraining existing models.

However, the models above often prioritize semantics, while acoustic features, independent of the semantic content, also play a significant role in speech. For example, features such as voice onset time (VOT) \cite{wu2020native}, formant frequencies \cite{feinberg2005manipulations}, and speaker-specific articulatory patterns \cite{wretling1998articulatory} are essential to maintain naturalness and prosodic consistency. 
Overemphasis on semantics can sometimes reduce attention to acoustic aspects,
affecting the overall consistency and prosodic detail.
Therefore, it is essential to process both the semantic and acoustic components in parallel when using LLMs for TTS, ensuring they are treated equally and generating semantic and acoustic tokens simultaneously.

To address these limitations, we first introduce a dual-layer synthesis framework that leverages a parallel AR model and a coupled NAR model to capture better the independence and interdependence of acoustic and semantic information. As shown in Figure \ref{1} (C), we first generate top parallel tokens that preserve the independence of semantic and acoustic features. Then, a coupled NAR model refines the synthesis by generating detailed tokens that capture their interdependence, resulting in comprehensive tokens.
Second, for speech token construction, we introduce the Parallel Tokenizer, which concurrently encodes both semantic and acoustic information. In LLM integration, we also use a Parallel Autoregressive Language Model to simultaneously generate semantic and acoustic tokens. These tokens are temporally aligned but content-wise independent, allowing the model to treat both aspects equally.

Our contributions can be summarized as follows:

\begin{itemize}

\item Parallel GPT presents a dual-layer synthesis framework harmonizing the Independence and Interdependence of acoustic and semantic Information. It consists of a parallel AR and coupled NAR model. The AR model generates the top tokens in parallel, maintaining the independence of acoustic and semantic features. The NAR model synthesizes the detailed tokens by jointly considering acoustic and semantic information, reflecting their interdependence.

\item Parallel GPT introduces the Parallel Tokenizer, which harmonizes the independence and interdependence of acoustic and semantic information. It simultaneously extracts both semantic and acoustic tokens, ensuring a balanced representation of speech.

\item Parallel GPT proposes a Parallel AR Language Model that generates both acoustic and semantic information concurrently. It predicts both types of tokens simultaneously, allowing the LLM to treat acoustic and semantic aspects equally.

\item We have conducted extensive experiments in both English and Chinese. We use the LibriTTS dataset for English and an internal dataset of 1062 hours of speech for Chinese. The results demonstrate that Parallel GPT significantly outperforms traditional zero-shot models’ synthesis quality and efficiency.
\end{itemize}

\section{Related Works}
This section describes related studies on speech token construction, LLM-based approaches for TTS, and AR-based and NAR-based TTS Models.

\subsection{Speech Token Construction}

Speech Token Construction refers to transforming speech into a latent space representation, enabling the reconstruction of the original speech. These representations can be discrete tokens or continuous vectors. Existing studies have explored this concept extensively \cite{defossez2022high} \cite{yang2023hifi} \cite{kumar2024high}.

Traditional approaches to speech token construction can be categorised into three primary types: Overall Speech Tokens, Only Semantic Tokens, and Two-Stage Tokens. Each of these methods represents a unique perspective on transforming speech into tokens. 

The Overall Speech Tokens method treats speech as an indivisible whole, coupling all the information into a unified tokenized representation. These models \cite{zeghidour2021soundstream} \cite{defossez2022high} \cite{yang2024simplespeech} focus on generating a single, comprehensive token stream. For example, methods like Encodec \cite{defossez2022high} compress speech into 8-bit tokens. It simplifies the modeling process, enhancing efficiency in storage and transmission while preserving essential speech characteristics.

The Only Semantic Tokens method focuses exclusively on extracting the linguistic content of speech, aiming to enhance its clarity and expressiveness. These models \cite{liu2024generative} \cite{kim22c_interspeech} \cite{yang2023uniaudio} emphasize semantic modeling and convert speech into semantic tokens. Additionally, many pre-trained SSL models for automatic speech recognition (ASR) simplify the token extraction process by reducing the training data required. For example, TransferTTS \cite{kim22c_interspeech} uses the
Wav2Vec \cite{schneider2019wav2vec}, and UnionAudio \cite{yang2023uniaudio} uses the HuBERT \cite{hsu2021hubert}.

The Two-Stage Tokens approach introduces a hierarchical framework by separating speech into lower-level semantic tokens and higher-level acoustic tokens. Approaches such as SpeechTokenizer \cite{zhang2024speechtokenizer} and MaskGCT \cite{wang2024maskgct} exemplify this method. 
They are not decoupled but rather extract speech at different levels.

The above three types of frameworks enormously improve zero-shot TTS performance. However, speech is inherently complex, and they do not fully capture the inherent independence between semantics and acoustics, which can lead to a loss of detail in speech synthesis. The Overall Speech Tokens are challenging in representing speech fully and can cause a loss of detailed information. The second and third kinds of tokens are emphasized first. There is a loss of the acoustic information independent of the semantic information, such as formant frequencies \cite{feinberg2005manipulations}, and speaker-specific articulatory patterns \cite{wretling1998articulatory}. 
Constructing speech tokens that are both semantically and acoustically decoupled yet synchronized is essential for harmonizing the independence and interdependence of acoustic and semantic information.

This paper proposes a novel Speech Token Construction method, the Parallel Tokenizer, which simultaneously generates semantic and acoustic tokens. It leverages different pre-trained SSL models for various downstream tasks \cite{baevski2020wav2vec} \cite{chen2022beats} \cite{wang2023cam++}. This approach also retains the advantages of using SSL models and reduces the training required for TTS data. 

\begin{figure*}[!t]
\centering
\includegraphics[width=0.9\textwidth]{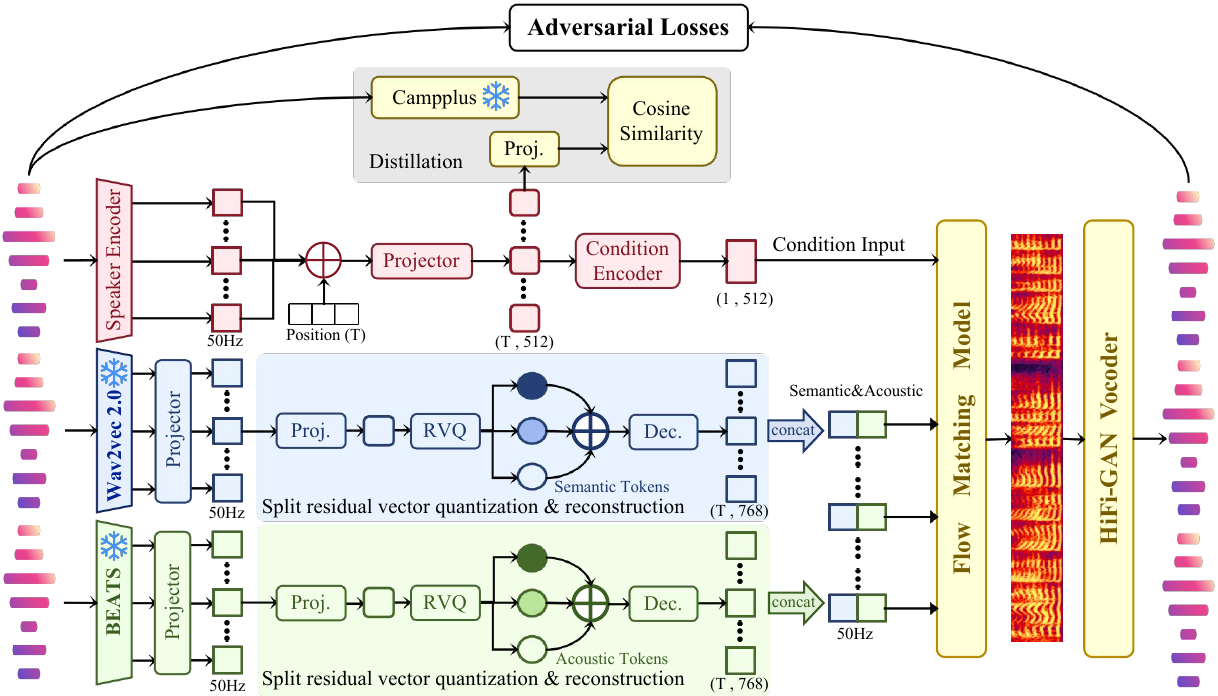}
\caption{The structure of Parallel Tokenizer. It is trained before training the TTS model. Wav2vec 2.0, BEATS, and Campplus models use downloaded pre-training weights and freeze the parameters during training.}
\label{3}
\end{figure*}

\subsection{LLM-based approaches for TTS}

Recent advancements in TTS have increasingly focused on utilizing LLMs to predict discrete speech tokens, offering a more flexible and expressive synthesis approach. Models like Transducer \cite{kim2023transduce}, UniAudio \cite{yang2023uniaudio}, Seed-TTS \cite{anastassiou2024seed}, and SoCodec \cite{guo2024socodec} demonstrate this shift. These models first extract semantic tokens from speech through pre-trained speech tokenizers. They then use these tokens to train LLMs, which predict speech patterns based on the underlying semantic information, transforming text into natural-sounding speech. This method improves naturalness, expressiveness, and adaptability, enabling zero-shot TTS across various speakers and domains.

However, speech is not solely driven by semantic information. Acoustic features are equally crucial. The models above prioritize semantic understanding but often fail to capture the full range of acoustic characteristics contributing to speech’s naturalness. Some models, such as Moshi \cite{defossez2024moshi} and Mini-Omni \cite{xie2024mini}, incorporate joint acoustic-semantic encoding to integrate acoustic features better.

Nevertheless, these models process semantic and acoustic features jointly rather than concurrently. The semantic layer dominates the loss function. Additionally, they typically fine-tune text-based LLMs, where the semantic aspect is prioritized. Focusing on semantics can sometimes lead to less emphasis on these acoustic aspects, which affects overall consistency and prosodic detail. Therefore, it is crucial to process both the semantic and acoustic components in parallel when using LLMs for TTS.

In this paper, we propose the Parallel Autoregressive Language Model, which processes temporally aligned semantic and acoustic tokens as inputs and simultaneously generates both types of tokens. This design ensures that semantic and acoustic information are treated equally, allowing the model to generate both token types in parallel. As a result, it significantly enhances the model’s ability to capture both aspects of speech, leading to more natural and expressive synthesis.

\subsection{AR-Based and NAR-Based TTS Models}

At a high modeling level, previous TTS models are usually categorized into AR-based and NAR-based models.

AR-based models, such as Tortoise-TTS \cite{betker2023better} and UniAudio \cite{yang2023uniaudio}, generate speech tokens sequentially using autoregressive methods. These models leverage AR models like LLaMA \cite{touvron2023llama} and GPT-2 \cite{radford2019language} to predict each token based on the preceding ones. The sequential generation framework enables AR models to produce speech that closely mirrors human-like patterns, delivering high-quality, expressive speech with remarkable naturalness.

NAR-based models, including E2-TTS \cite{eskimez2024e2} and F5-TTS \cite{chen2024f5}, utilize advanced techniques like Diffusion \cite{ho2020denoising}, Normalized flow \cite{papamakarios2021normalizing}, and Flow-matching \cite{lipman2022flow} models for speech synthesis. These models bypass the sequential generation framework. They enable faster and more robust synthesis.

However, both AR and NAR models have limitations in speech synthesis. AR models face challenges when predicting complex tokens. NAR models, in contrast, struggle with duration prediction. Incorrect duration estimates cause timing mismatches, resulting in unnatural speech.

This paper proposes splitting parallel tokens into top and detailed tokens \cite{barnes1996advances}. The AR model predicts the top tokens, while the NAR model refines the details by predicting the detailed tokens based on the top tokens. This approach leverages the simplicity of AR prediction, with the NAR model using AR output to resolve duration prediction issues, ensuring more accurate timing and natural speech. Unlike VALL-E \cite{wang2023neural}, our approach incorporates speech characteristics in both semantic and acoustic token prediction. The AR module emphasizes their independence by separately predicting semantic and acoustic features. In contrast, the NAR module captures their interdependence by jointly considering both types of information, refining the prediction of detailed features in a coupled manner.

\section{Proposd Method}

In this part, we present the details of the proposed Parallel GPT. The overall architecture of the Parallel GPT inference pipeline is documented in Figure \ref{2}, which includes three new parts: (1) A Parallel Tokenizer to extract acoustic and semantic tokens from speech and generate continuous features from parallel tokens; (2) A Parallel Autoregressive Language Model; (3) A Coupled Non-Autoregressive Transformer. In addition to the parts above, the text-to-phoneme conversion is performed using a dictionary lookup approach. The speech vocoder is based on HiFi-GAN \cite{kong2020hifi}.

In the following section, we first introduce the Parallel Tokenizer, Parallel Autoregressive Language Model, and Coupled Non-Autoregressive Transformer, and then we provide details of the model training.
\subsection{Parallel Tokenizer}
The Parallel Tokenizer is a pivotal component of the Parallel GPT system. It converts complex speech into parallel tokens that distinctly capture semantic and acoustic aspects. These tokens can reconstruct high-quality speech incorporating speaker-specific information. The architecture of the Parallel Tokenizer is illustrated in Figure \ref{3}.

\begin{figure*}[!t]
\centering
\includegraphics[width=0.9\textwidth]{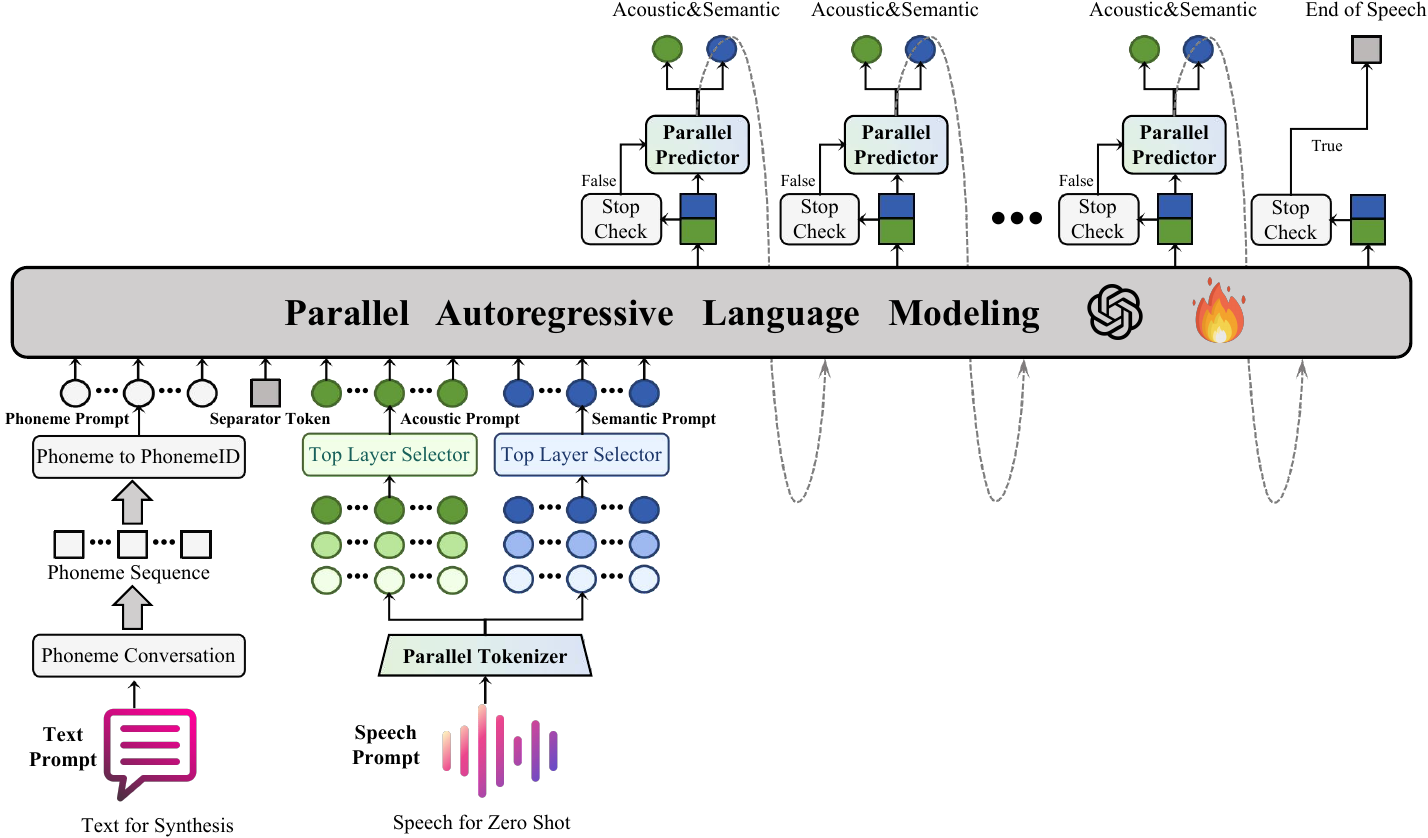}
\caption{The structure of the Parallel Autoregressive Language Model. It is designed to synthesize target top tokens from the target text and reference speech. We define the first dimension of the tokens obtained by the Parallel Tokenizer as the "top tokens", representing the most important elements.}
\label{4}
\end{figure*}

Recognizing that different SSL models are specialized for various downstream tasks, the Parallel Tokenizer combines the strengths of pre-trained SSL models to generate comprehensive representations. Specifically, BEATS \cite{chen2022beats}, an audio encoder, extracts acoustic features. Wav2Vec 2.0 \cite{baevski2020wav2vec}, a widely adopted SSL model trained for masked reconstruction tasks in ASR, is used to extract semantic features. Campplus \cite{wang2023cam++}, an SSL model designed for speaker recognition, is also leveraged to extract speaker-specific information.

In extracting acoustic and semantic tokens, pre-trained Wav2Vec 2.0 and BEATS project 16 kHz speech into 768-dimensional embeddings with a 50 Hz sampling rate. Both models employ a projector to adjust the internal temporal alignment, optimizing content-level alignment. This temporal adjustment is non-causal and compatible with flow-based inference, as these embeddings are only utilized during training. Subsequently, the 768-dimensional linear projections are passed through the Residual Vector Quantization (RVQ) encoder, generating 3-bit token outputs respectively. These tokens are then decoded back into 768-dimensional features using the RVQ decoder.

Regarding speaker information, the audio is first converted into 512-dimensional features sampled at 50 Hz through a speaker encoder. The speaker encoder follows the same architecture as ECAPA-TDNN \cite{desplanques20_interspeech} but retains the temporal dimension of the speaker embedding. It is crucial since speaker characteristics are not limited to timbre but also involve acoustic details. The encoder output is augmented with position encoding and passed through a projector to capture temporal dependencies. A pre-trained Campplus model is employed to extract a 192-dimensional speaker feature. We compute the cosine distance between the extracted speaker features and the transformed 512-dimensional features, performing distillation to align the speaker’s characteristics. Finally, the speaker feature is processed by a condition encoder, transforming it into a 512-dimensional time-invariant embedding. The condition encoder comprises an initial convolutional layer and several multi-head attention blocks. It has the same structure as the condition encoder in Tortoise-TTS \cite{betker2023better}.

For speech reconstruction, the semantic and acoustic features output by the RVQ decoder are concatenated in the temporal dimension. These concatenated features are fed into the Flow Matching Model, with the speaker representations as the conditional input. The generated Mel spectrogram is subsequently input into the Hifi-GAN \cite{kong2020hifi} vocoder for speech synthesis. For adversarial training, a multi-scale discriminator and a multi-period discriminator are employed. The goal is to reconstruct and match the speech to the ground truth. The configurations of these discriminators are consistent with those used in Hifi-GAN.

The loss function for training the Parallel Tokenizer, $\mathcal{L}_{\text{PT}}$, is defined as:
\begin{equation}
\mathcal{L}_{\text{PT}} = \mathcal{L}_{\text{Sem}} + \mathcal{L}_{\text{Acous}} + \mathcal{L}_{\text{Speaker}} + \mathcal{L}_{\text{Mel}} + \mathcal{L}_{\text{Adv}}
\end{equation}
where:
\begin{itemize}
    \item $\mathcal{L}_{\text{Sem}}$ is the semantic RVQ reconstruction loss, computed as the mean squared error (MSE) between the reconstructed semantic features (after RVQ encoding and decoding) and the original semantic features.
    \item $\mathcal{L}_{\text{Acous}}$ is the acoustic RVQ reconstruction loss, computed as the MSE between the reconstructed acoustic features (after RVQ encoding and decoding) and the original acoustic features.
    \item $\mathcal{L}_{\text{speaker}}$ is the cross-entropy loss between the predicted speaker representation and the ground truth speaker representation extracted by the pre-trained CampPlus model.
    \item $\mathcal{L}_{\text{Mel}}$ is the MSE loss between the Mel spectrogram generated by the Flow Matching Model and the ground truth Mel spectrogram.
    \item $\mathcal{L}_{\text{Adv}}$ is the adversarial loss computed between the reconstructed speech and the ground truth speech, using the multi-scale and multi-period discriminators.
\end{itemize}

\subsection{Parallel Autoregressive Language Model}

The structure of the Parallel Autoregressive Language Model is shown in Figure \ref{4}. It takes input text and reference speech to generate the top parallel tokens for synthesizing speech simultaneously.

This module is based on a double-layer GPT-2 \cite{radford2019language} decoder-only Transformer. We first extract the reference parallel tokens from the reference speech during the generation process using the pre-trained Parallel Tokenizer. The parallel tokens are divided into semantic and acoustic tokens with three encoding layers. 
For the Parallel Tokenizer, since RVQ is employed, the first quantizer plays the most crucial role in reconstruction, with the influence of the other quantizers gradually decreasing. Therefore, the first layer is selected as the top tokens. The reference speech’s top semantic and acoustic tokens and text are projected into an embedding space with the exact dimensions. All streams of the speech sequence are processed separately and then combined. The text and speech sequences are augmented with different learnable positional embeddings, while the acoustic and semantic tokens of the speech share the same set of learnable positional embeddings. This embedding sequence is then processed through a stack of causal Transformer layers.

The model's embedding output is 1024-dimensional, split into two 512-dimensional vectors at each autoregressive step. These two vectors are simultaneously passed to a stop check module to determine whether to stop the sequence generation.
If the stop check module decides to stop (based on these vectors), the model outputs the "end of sequence" (EOS) token. If the stop check module decides to continue (based on these vectors), both 512-dimensional outputs are passed to the parallel predictor. The parallel predictor consists of two independent logic predictors, each predicting the probabilities for semantic and acoustic tokens, using linear layers. An autoregressive (AR) sampling strategy with top-k selection is then applied to determine the final output token from the semantic and acoustic predictions.

During the prediction process, the top semantic and acoustic tokens are split along the dimensional axis and processed by different logic predictors. 
The semantic and acoustic tokens are independently generated in content but aligned in timing and duration. Rather than two separate AR models, a parallel AR model is used to predict the top token for each token type (semantic and acoustic).

The loss function for training the Parallel Autoregressive Language Model, \( \mathcal{L}_{\text{PAR}} \), is defined as:

\begin{equation}
\mathcal{L}_{\text{PAR}} = \mathcal{L}_{\text{semantic}} + \mathcal{L}_{\text{acoustic}} + \mathcal{L}_{\text{stop}}
\end{equation}

Where:

\begin{itemize}
    \item \( \mathcal{L}_{\text{semantic}} \) is the semantic prediction loss, computed as the cross-entropy loss between the predicted semantic tokens and the ground truth semantic tokens:
    \begin{equation}
    \mathcal{L}_{\text{semantic}} = -\sum_{t=1}^{T} \log P(\hat{s}_t | s_{<t})
    \end{equation}
    Here, \( \hat{s}_t \) represents the predicted semantic token at time step \( t \), and \( s_{<t} \) represents all the previous semantic tokens. \( P(\hat{s}_t | s_{<t}) \) is the probability of predicting \( \hat{s}_t \) given the preceding semantic tokens.

    \item \( \mathcal{L}_{\text{acoustic}} \) is the acoustic prediction loss, computed as the cross-entropy loss between the predicted acoustic tokens and the ground truth acoustic tokens:
    \begin{equation}
    \mathcal{L}_{\text{acoustic}} = -\sum_{t=1}^{T} \log P(\hat{a}_t | a_{<t})
    \end{equation}
    Here, \( \hat{a}_t \) represents the predicted acoustic token at time step \( t \), and \( a_{<t} \) represents all the previous acoustic tokens. \( P(\hat{a}_t | a_{<t}) \) is the probability of predicting \( \hat{a}_t \) given the preceding acoustic tokens.
    \item \( \mathcal{L}_{\text{stop}} \) is the stop-check loss. It is computed as the cross-entropy loss between the predicted stop token probability and the ground truth stop decision, where the semantic and acoustic tokens jointly determine the stop decision:
    \begin{equation}
    \mathcal{L}_{\text{stop}} = -\log P(\text{STOP} | s_T, a_T)
    \end{equation}
    Here, \( s_T \) and \( a_T \) represent the semantic and acoustic tokens with the same time step, and \( P(\text{STOP} | s_T, a_T) \) is the probability that the sequence should stop based on these tokens.
\end{itemize}

\begin{figure}[!t]
\centering
\includegraphics[width=\columnwidth]{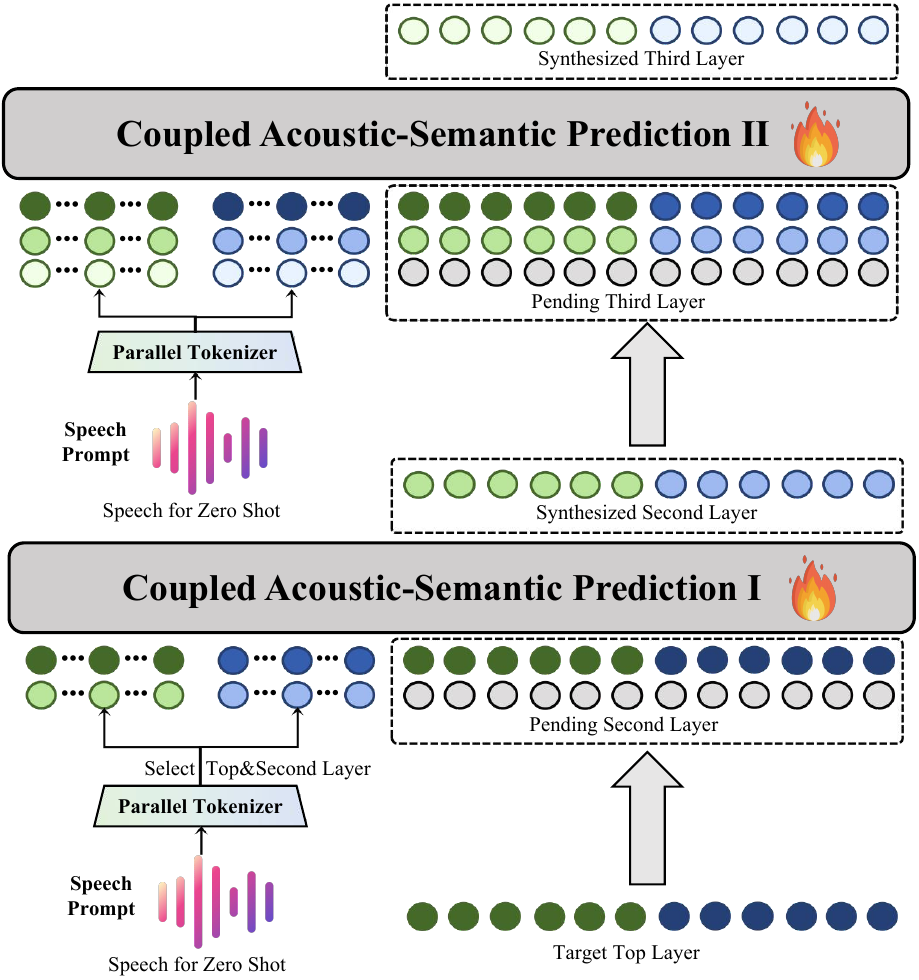}
\caption{The structure of Coupled Non-Autoregressive Transformer. It is designed to generate detailed tokens from target top tokens and reference speech.}
\label{5}
\end{figure}

\subsection{Coupled Non-Autoregressive Transformer}

The Coupled Non-Autoregressive Transformer is designed to generate detailed tokens from the top parallel tokens and reference speech. Its structure is shown in Figure \ref{5}.

After obtaining the top parallel tokens through the Parallel Autoregressive Language Model, the NAR model generates the remaining two-layer tokens for both the semantic and acoustic aspects. 
Considering that the influence of the other quantizers gradually decreases in the RVQ encoder’s output, the prediction in this module is divided into two steps. First, the second-layer tokens are predicted using the top tokens. In the second step, the third-layer tokens are predicted using both the top and second-layer tokens.

The semantic and acoustic tokens are mapped together and not split along the dimensions. This is because the subsequent prediction needs to account for the interdependence between semantics and acoustics in speech rather than treating them independently, as in the parallel AR model.
The two Coupled Acoustic-Semantic Prediction layers comprise a three-layer Transformer decoder and a 1024-dimensional classifier. The output is then mapped to the number of token categories.
Multiple self-attention layers in the model capture the coupled relationship between semantic and acoustic tokens. The detailed tokens are predicted by leveraging the inherent relationships between the top acoustic and semantic tokens.

The loss function for training the Coupled Acoustic-Semantic Prediction layer, \( \mathcal{L}_{\text{Coupled}} \), is defined as:

\begin{equation}
\mathcal{L}_{\text{Coupled}} = \mathcal{L}_{\text{second}} + \mathcal{L}_{\text{third}}
\end{equation}

Where:

\begin{itemize}
    \item \( \mathcal{L}_{\text{second}} \) is the loss for predicting the second-layer tokens, computed as the cross-entropy loss between the predicted second-layer tokens and the ground truth second-layer tokens. The condition includes the top token and the first and second-layer tokens from the reference speech, as extracted by the pre-trained Parallel Tokenizer:
    \begin{equation}
    \mathcal{L}_{\text{second}} = - \sum_{t=1}^{T} \log P(\hat{c}_2 | c_1, c_{\text{ref1}}, c_{\text{ref2}})
    \end{equation}
    Here, \( \hat{c}_2 \) represents the predicted second-layer token, \( c_1 \) is the top token, and \( c_{\text{ref1}} \) and \( c_{\text{ref2}} \) are the first and second-layer tokens from the reference speech, respectively. These reference tokens are obtained by passing the reference speech through a pre-trained Parallel Tokenizer. The loss is computed as the cross-entropy between the predicted second-layer token and the ground truth token.

    \item \( \mathcal{L}_{\text{third}} \) is the loss for predicting the third-layer tokens, computed as the cross-entropy loss between the predicted third-layer tokens and the ground truth third-layer tokens. The condition includes the top token, the second-layer token, and the first, second, and third-layer tokens from the reference speech:
    \begin{equation}
    \mathcal{L}_{\text{third}} = - \sum_{t=1}^{T} \log P(\hat{c}_3 | c_1, c_2, c_{\text{ref1}}, c_{\text{ref2}}, c_{\text{ref3}})
    \end{equation}
    Here, \( \hat{c}_3 \) represents the predicted third-layer token, \( c_1 \) and \( c_2 \) are the top and second-layer tokens, respectively, and \( c_{\text{ref1}}, c_{\text{ref2}}, c_{\text{ref3}} \) are the first, second, and third-layer tokens from the reference speech. The pre-trained Parallel Tokenizer also extracts these reference tokens. The loss is computed as cross-entropy between the predicted third layer and ground truth tokens.

    \item  In each loss calculation, the semantic and acoustic tokens are considered jointly without separation.

\end{itemize}

\section{Experimental Setup}
\subsection{Dataset}
\subsubsection{English Speech Dataset}
The LibriTTS \cite{zen2019libritts} dataset is used. All comparison and ablation experiments are conducted at a sampling rate of 16 kHz.

LibriTTS is a large-scale multi-speaker corpus for English speech synthesis, containing 585 hours of speech data from more than 2,300 speakers. We divide the LibriTTS dataset into three parts: the training, development, and test sets. The training set is constructed by merging three subsets: train-clean-100, train-clean-360, and train-other-500. The development set is created by combining the dev-clean and dev-other subsets, while the test set is formed by merging the test-clean and test-other subsets. The training set is used for model training, while the development and test sets are used for evaluation. The development set includes speakers from the training set, while the test set contains speakers not seen during training, allowing for the assessment of the model’s performance on both seen and unseen speakers.

\subsubsection{Chinese Speech Dataset}
An internal Chinese dataset is used. It consists of read speech from Chinese novels with a sampling rate of 16 kHz. The dataset includes 453,716 audio samples, totaling 1,062 hours of speech recorded by 43,034 unique speakers. Two speech cleaning tools, Emilia pipeline \cite{emilia} and NCSSD pipeline \cite{liu2024generative}, are applied for preprocessing and noise reduction to ensure high-quality data.

The dataset is divided into training and test sets. Audio samples from 42,000 speakers are used for training, while the remaining speakers are allocated to the test set. This split ensures that the model is trained on a broad range of speakers, with the test set evaluating the model’s ability to generalize to unseen speakers. The dataset is employed to train and assess the performance of speech synthesis models for Chinese speech.

\subsection{Model Training and Inference}
\subsubsection{Parallel Tokenizer}

For the training of the Parallel Tokenizer, we first download the pre-trained model weights of Wav2Vec 2.0\footnote{ Wav2Vec 2.0 Base: https://github.com/eastonYi/wav2vec}, BEATs\footnote{BEATs Iter3+: https://github.com/microsoft/unilm/tree/master/beats}, and Campplus\footnote{Campplus: https://www.modelscope.cn/models/iic/CosyVoice-300M/file/ view/master?fileName=campplus.onnx}. These weights are then placed into the corresponding components of the Parallel Tokenizer, with the weights frozen during training. The model is trained separately on Chinese and English datasets, with each language trained individually. The AdamW optimizer is used with an initial learning rate of \( 2.0 \times 10^{-4} \), which decays by a factor of \( 0.999^{1/8} \) every epoch.
We train the Parallel Tokenizer models with 450k steps.

\subsubsection{Parallel GPT}

To train the Parallel GPT model, we freeze the Parallel Tokenizer. We train the parallel autoregressive language and coupled non-autoregressive transformer models separately. The parallel autoregressive language model is trained to predict the target top tokens from text and prompts. The coupled non-autoregressive transformer model is trained to predict top tokens from detailed tokens and to convert the comprehensive tokens into speech.

The parallel autoregressive language model is trained for 800K steps on an NVIDIA A800 GPU with a batch size of 16. During training, the learning rate is initially set to \( 1.0 \times 10^{-2} \), with a warm-up period of 2k steps. After the warm-up, the learning rate decays according to a cosine schedule over the next 40K steps, starting at \( 1.0 \times 10^{-5} \) and ending at \( 1.0 \times 10^{-4} \). The optimizer used is ScaledAdam, with \( \beta_1 = 0.9 \), \( \beta_2 = 0.95 \), and a gradient clipping scale of 2.0, applied every 1k steps.

The coupled non-autoregressive transformer model is trained separately for 200K steps with a batch size of 16. It uses a fixed learning rate of \( 2.0 \times 10^{-5} \) and the AdamW optimizer. Additionally, the model incorporates the pre-trained decoder weights from the Parallel Tokenizer and fine-tunes them. The focus is predicting detailed tokens from top tokens and converting comprehensive tokens into speech.

\subsection{Baselines}

To thoroughly assess the improvements introduced by our approach, we design a comprehensive set of comparison experiments. These experiments span many TTS models, including those based on pre-trained SSL models. We compare models that utilize speech tokens, as well as those that incorporate both acoustic and semantic tokens, highlighting the advantages of our method. Additionally, we evaluate both AR and NAR models to demonstrate the versatility and robustness of our approach. This carefully constructed experimental setup allows us to effectively assess the performance of our model in various contexts, ensuring that our contributions lead to meaningful improvements in speech synthesis quality.

\subsubsection{TTS baselines on English Speech Synthesis}

We conduct comprehensive experiments to evaluate English speech synthesis, retraining, or fine-tuning the models on the same training datasets. Each model represents a distinct approach that contributes to a thorough comparison. The models included in our comparison, along with their corresponding code and model sources, are listed below.

\begin{itemize}
\item \textbf{YourTTS}\footnote{https://github.com/Edresson/YourTTS} \cite{casanova2022yourtts}: A non-autoregressive TTS model designed for zero-shot speech synthesis, generating speech from text and reference audio.

\item \textbf{TransferTTS}\footnote{https://github.com/hcy71o/TransferTTS} \cite{kim22c_interspeech}: A non-autoregressive TTS model for zero-shot speech synthesis, leveraging a pre-trained SSL model, Wav2Vec \cite{schneider2019wav2vec}, for speech representation extraction.

\item \textbf{VALL-E}\footnote{https://github.com/lifeiteng/VALL-E} \cite{wang2023neural}: An autoregressive TTS model that uses pre-trained Encodec \cite{defossez2022high} for the extraction of speech tokens.

\item \textbf{UniAudio}\footnote{https://github.com/yangdongchao/UniAudio} \cite{yang2023uniaudio}: A universal autoregressive audio generation model evaluated for its zero-shot TTS capability employing pre-trained SSL models like HuBERT \cite{hsu2021hubert} for speech representation.

\item \textbf{E2-TTS}\footnote{https://github.com/lucidrains/e2-tts-pytorch} \cite{eskimez2024e2}: A fully non-autoregressive zero-shot TTS system based on character sequences and padding tokens.

\item \textbf{CosyVoice}\footnote{https://github.com/FunAudioLLM/CosyVoice} \cite{du2024cosyvoice}: An autoregressive zero-shot TTS model based on supervised semantic tokens (FunCodec \cite{du2024funcodec}), combining large language models with conditional flow matching for speech synthesis.

\item \textbf{MaskGCT}\footnote{https://github.com/open-mmlab/Amphion/blob/main/models/tts/maskgct} \cite{wang2024maskgct}: A two-stage, fully non-autoregressive zero-shot text-to-speech model. It uses W2v-BERT 2.0 \cite{chung2021w2v} to extract semantic features and SpeechTokenizer’s training method to generate semantic and acoustic tokens. It synthesizes semantic tokens first and then predicts acoustic tokens from them to generate speech.

\end{itemize}

\subsubsection{TTS baselines on Chinese Speech Synthesis}

We select several state-of-the-art models, including CosyVoice, E2-TTS, and MaskGCT, as baselines for comparing Chinese speech synthesis. These models represent a range of autoregressive and non-autoregressive approaches. All models are retrained or fine-tuned on the same dataset to ensure a fair and direct performance comparison.

\renewcommand{\arraystretch}{1.1}
\begin{table*}[!t]
\caption{Comparative Experiments on LibriTTS Dataset for English Zero-Shot Text-to-Speech. SBS refers to SpeechBERTScore.}
\label{english}
\centering
\resizebox{1\textwidth}{!}{ 
\begin{tabular}{|c|c|c|c|c|c|c|c|c|c|c|}
\hline
\cline{2-11}
& \multicolumn{5}{c|}{\textbf{LibriTTS Development Set}} & \multicolumn{5}{c|}{\textbf{LibriTTS Test Set}} \\
\cline{2-11}
\textbf{Model Name} & \multicolumn{2}{c|}{\textbf{Subjective}} & \multicolumn{3}{c|}{\textbf{Objective}} & \multicolumn{2}{c|}{\textbf{Subjective}} & \multicolumn{3}{c|}{\textbf{Objective}} \\
\cline{2-11}
& \hspace{1em}\textbf{MOS ($\uparrow$)}\hspace{1em} & \hspace{1em}\textbf{SMOS ($\uparrow$)}\hspace{1em} & \textbf{UTMOS ($\uparrow$)} & \textbf{WER ($\downarrow$)} & \textbf{SBS ($\uparrow$)} & \hspace{1em}\textbf{MOS ($\uparrow$)}\hspace{1em} & \hspace{1em}\textbf{SMOS ($\uparrow$)}\hspace{1em} & \textbf{UTMOS ($\uparrow$)} & \textbf{WER ($\downarrow$)} &\textbf{SBS ($\uparrow$)} \\
\hline
\hline
Ground Truth &  4.76 $\pm$ \scriptsize{0.08}  & 4.79$\pm$ \scriptsize{0.06} & 4.086 & 0.106 & 1.000 & 4.75$\pm$ \scriptsize{0.08} & 4.65$\pm$ \scriptsize{0.06} & 3.929 & 0.161 & 1.000 \\
\hline
YourTTS \cite{casanova2022yourtts} & 
3.65 $\pm$ \scriptsize{0.10}& 3.65 $\pm$ \scriptsize{0.09} & 3.357 & 0.234 & 0.788 & 3.58$\pm$ \scriptsize{0.09} & 3.51$\pm$ \scriptsize{0.09} & 3.404 & 0.284 & 0.781\\

TransferTTS \cite{kim22c_interspeech} & 3.77$\pm$ \scriptsize{0.09} & 3.75$\pm$ \scriptsize{0.09} & 3.358 & 0.237 & 0.830 & 3.63$\pm$ \scriptsize{0.10} & 3.50$\pm$ \scriptsize{0.09} & 3.506 & 0.242 & 0.821 \\

VALL-E \cite{wang2023neural} & 3.57$\pm$ \scriptsize{0.10} & 3.43$\pm$ \scriptsize{0.08} & 3.001 & 0.304 & 0.798 & 3.35$\pm$ \scriptsize{0.10} & 3.31$\pm$ \scriptsize{0.09} & 2.980 & 0.432 & 0.695 \\

UniAudio \cite{yang2023uniaudio} & 3.75$\pm$ \scriptsize{0.09} & 3.63$\pm$ \scriptsize{0.10} & 3.259 & 0.285 & 0.800 & 3.77$\pm$ \scriptsize{0.09} & 3.70$\pm$ \scriptsize{0.10} & 3.129 & 0.317 & 0.712 \\

E2-TTS \cite{eskimez2024e2} & 3.52$\pm$ \scriptsize{0.09} & 3.87$\pm$ \scriptsize{0.01} & 3.085 & 0.284 & 0.829 & 3.52$\pm$ \scriptsize{0.09} & 3.78$\pm$ \scriptsize{0.10} & 3.002 & 0.257 & 0.818 \\

CosyVoice \cite{du2024cosyvoice} & \underline{4.05$\pm$ \scriptsize{0.10}} & \textbf{4.23$\pm$ \scriptsize{0.10}} & \textbf{3.903} & 0.340 & \underline{0.823} & 4.01$\pm$ \scriptsize{0.11} & \textbf{4.11$\pm$ \scriptsize{0.11}} & \textbf{3.803} & 0.303 & 0.816 \\

MaskGCT \cite{wang2024maskgct} & 4.01$\pm$ \scriptsize{0.10} & 4.05$\pm$ \scriptsize{0.10} & 3.600 & \underline{0.216} & 0.821 & 4.02$\pm$ \scriptsize{0.10} & 3.91$\pm$ \scriptsize{0.11} & 3.520 & \underline{0.251} & \underline{0.819} \\

Parallel GPT & \textbf{4.11$\pm$ \scriptsize{0.09}} & \underline{4.08$\pm$ \scriptsize{0.10}} & \underline{3.632} & \textbf{0.211} & \textbf{0.824} & \textbf{4.08$\pm$ \scriptsize{0.13}} & \underline{3.92$\pm$ \scriptsize{0.11}} & \underline{3.642} & \textbf{0.241} & \textbf{0.825} \\
\hline
\end{tabular}
}
\end{table*}

\begin{table*}[!t]
\caption{Comparative Experiments on  Chinese Internal Dataset for Chinese Zero-Shot Text-to-Speech. SBS refers to SpeechBERTScore.}
\label{chinese}
\centering
\resizebox{1\textwidth}{!}{ 
\begin{tabular}{|c|c|c|c|c|c|c|c|c|c|c|}
\hline
\cline{2-11}
& \multicolumn{5}{c|}{\textbf{Chinese Development Set}} & \multicolumn{5}{c|}{\textbf{Chinese Test Set}} \\
\cline{2-11}
\textbf{Model Name} & \multicolumn{2}{c|}{\textbf{Subjective}} & \multicolumn{3}{c|}{\textbf{Objective}} & \multicolumn{2}{c|}{\textbf{Subjective}} & \multicolumn{3}{c|}{\textbf{Objective}} \\
\cline{2-11}
& \hspace{1em}\textbf{MOS ($\uparrow$)}\hspace{1em} & \hspace{1em}\textbf{SMOS ($\uparrow$)}\hspace{1em} & \textbf{UTMOS ($\uparrow$)} & \textbf{WER ($\downarrow$)} & \textbf{SBS ($\uparrow$)} & \hspace{1em}\textbf{MOS ($\uparrow$)}\hspace{1em} & \hspace{1em}\textbf{SMOS ($\uparrow$)}\hspace{1em} & \textbf{UTMOS ($\uparrow$)} & \textbf{WER ($\downarrow$)} &\textbf{SBS ($\uparrow$)} \\
\hline
\hline
Ground Truth & 4.56$\pm$ \scriptsize{0.08} & 4.66$\pm$ \scriptsize{0.07} &  {---} & 0.080 & 1.000 & 4.78$\pm$ \scriptsize{0.06} & 4.81$\pm$ \scriptsize{0.06} & {---}& 0.118 & 1.000 \\
\hline
CosyVoice \cite{du2024cosyvoice} & \underline{3.99$\pm$ \scriptsize{0.10}} & \underline{4.23$\pm$ \scriptsize{0.10}} & {---}& 0.198 & \underline{0.798} & \underline{4.13$\pm$ \scriptsize{0.11}} & \textbf{4.28$\pm$ \scriptsize{0.10}} & {---} & 0.214 & \underline{0.821} \\

E2-TTS \cite{wang2024maskgct} & 3.98$\pm$ \scriptsize{0.09} & 4.16$\pm$ \scriptsize{0.10} &{---} & \underline{0.195} & 0.766 & 3.88$\pm$ \scriptsize{0.11} & 4.06$\pm$ \scriptsize{0.10} & {---} & \underline{0.198} & 0.738 \\

MaskGCT \cite{eskimez2024e2} & 3.86$\pm$ \scriptsize{0.11} & 4.00$\pm$ \scriptsize{0.10} & {---} & 0.211 & 0.788 & 4.05$\pm$ \scriptsize{0.11} & 4.12$\pm$ \scriptsize{0.11} & {---} & 0.199 & 0.820 \\

Parallel GPT & \textbf{4.23$\pm$ \scriptsize{0.09}} & \textbf{4.26$\pm$ \scriptsize{0.09}} & {---} & \textbf{0.185} & \textbf{0.799} & \textbf{4.19$\pm$ \scriptsize{0.09}} & \underline{4.19$\pm$ \scriptsize{0.10}} & {---} & \textbf{0.193} & \textbf{0.826} \\
\hline
\end{tabular}
}
\end{table*}

\subsection{Evaluation Metrics}

We perform objective evaluations in multiple dimensions for the TTS task with zero-shot, including semantic similarity, speech quality, and speaker similarity. Additionally, subjective evaluations are conducted to provide a more nuanced assessment of the synthesis performance.

\subsubsection{Objective Evaluation}

We conduct objective experiments to evaluate the performance of the model using three key metrics: SpeechBERTScore \cite{saeki2024speechbertscore}, UTMOS \cite{saeki2022utmos}, and Word Error Rate (WER). SpeechBERTScore provides a reliable measure of the semantic similarity between synthesized and actual speech, evaluating the extent to which the generated speech conveys the intended meaning. UTMOS assesses the overall naturalness and quality of the generated speech by comparing it with manual judgment results, serving as a valid measure of perceived quality. For WER evaluation, we use a Wav2Vec 2.0-large-based automatic speech recognition system to measure speech recognition accuracy, providing an objective indicator of transcription quality.

\subsubsection{Subjective Evaluation}
In the subjective evaluation, 20 human evaluators rate a set of randomly selected samples, providing separate scores for naturalness and similarity. Each evaluator gives a score on a 5-point scale from 1 to 5 for both aspects.

We conduct a Similarity Mean Opinion Score (SMOS) test to evaluate the speaker similarity between the synthesized speech and the reference speech. Additionally, the Mean Opinion Score (MOS) assesses the quality and naturalness of the generated speech.
For the SMOS test, we ask evaluators to assess the similarity of the generated speech to the reference speech in terms of timbre, emotion, and prosody.
For the MOS test, we ask evaluators to evaluate the speech’s realism, naturalness, clarity, and prosody.

\section{Experiments}

In this section, we evaluate the performance of Parallel GPT in a zero-shot speech synthesis task. It includes comparison, ablation, and unique experiments for decoupled characterization of Parallel Tokens. The corresponding demos are provided in the abstract, and we encourage readers to refer to them to better understand the results.

\subsection{Experiments on Zero-Shot Text-to-Speech}

We assess Parallel GPT's performance on zero-shot speech synthesis through comprehensive experiments conducted on both Chinese and English datasets. The English experimental results are detailed in Table \ref{english}, with corresponding Chinese results presented in Table \ref{chinese}. Both SMOS and MOS scores are reported with 95\% confidence intervals. Our analysis focuses on two key aspects: speech quality and speaker similarity.

\subsubsection{Speech Quality Evaluation}

Speech quality serves as a fundamental metric for TTS systems, directly determining the naturalness and intelligibility of synthesized speech. Parallel GPT significantly enhances speech naturalness by effectively integrating semantic and acoustic information through its parallel AR and NAR frameworks. Our experimental results demonstrate that Parallel GPT achieves superior MOS and UTMOS scores compared to traditional models in both English and Chinese tests, highlighting its exceptional capability in generating natural, high-quality speech. Furthermore, the WER metric indicates notable improvements in semantic accuracy and speech clarity. While these results are promising, we acknowledge there remains potential for enhancement when compared to cutting-edge models in the field.

\subsubsection{Speaker Similarity Assessment}

Speaker similarity evaluation focuses on three crucial dimensions: timbre matching, prosody consistency, and semantic alignment between synthesized and target speaker voices. Through SMOS and SpeechBERTScore metrics, our experiments reveal that Parallel GPT achieves good timbre and prosody reproduction performance, surpassing several benchmark models in SMOS evaluations. The synthesized speech exhibits remarkable fidelity to the target speaker's vocal characteristics. Additionally, SpeechBERTScore analysis confirms high semantic similarity, with Parallel GPT maintaining consistent performance across both English and Chinese language datasets.

\subsubsection{Results Analysis}

Experimental results demonstrate that Parallel GPT outperforms CosyVoice, E2TTS, and MaskGCT in both speech naturalness (MOS) and semantic-acoustic alignment (WER/SpeechBERTScore). Unlike traditional cascaded TTS systems, Parallel GPT's parallel semantic-acoustic modeling optimizes discrete linguistic and continuous speech representations, improving prosody. While conventional approaches process semantics and acoustics sequentially, resulting in information loss, our parallel generation strategy enables dynamic coordination between the two modalities, producing more natural intonation and stress patterns that better reflect semantic intent.

Regarding semantic fidelity, Parallel GPT achieves significantly lower WER and higher SpeechBERTScore, demonstrating its ability to bridge the representation gap between text and speech. Traditional pipelines often introduce pronunciation ambiguities or prosodic inconsistencies due to the disconnected text encoding and acoustic generation stages. In contrast, our parallel modeling allows the LLM to condition acoustic token generation on real-time semantic context, thereby improving articulation accuracy and semantic coherence.

Although Parallel GPT exhibits a minor deficit in speaker similarity (SMOS) compared to CosyVoice (gap <0.2 MOS), this difference is negligible. CosyVoice's superior SMOS stems from its specialized speaker embedding refinement during token-to-speech conversion, while our model currently prioritizes joint semantic-acoustic generation. This suggests that future improvements in speaker modeling could close this gap without compromising Parallel GPT's existing advantages.

In conclusion, Parallel GPT's parallel generation architecture validates the superiority of joint semantic-acoustic modeling, particularly in prosody and semantic fidelity. Subsequent work may focus on speaker-specific refinements to achieve comprehensive improvements over current state-of-the-art systems.

\begin{table*}[!t]
\caption{Ablation Experiments on English Datasets. SBS refers to SpeechBERTScore.}
\label{ablation}
\centering
\resizebox{1\textwidth}{!}{ 
\begin{tabular}{|l|c|c|c|c|c|c|c|c|c|c|}
\hline
\cline{2-11}
& \multicolumn{5}{c|}{\textbf{LibriTTS Development Set}} & \multicolumn{5}{c|}{\textbf{LibriTTS Test Set}} \\
\cline{2-11}
\textbf{Model Name} & \multicolumn{2}{c|}{\textbf{Subjective}} & \multicolumn{3}{c|}{\textbf{Objective}} & \multicolumn{2}{c|}{\textbf{Subjective}} & \multicolumn{3}{c|}{\textbf{Objective}} \\
\cline{2-11}
& \hspace{1em}\textbf{MOS ($\uparrow$)}\hspace{1em} & \hspace{1em}\textbf{SMOS ($\uparrow$)}\hspace{1em} & \textbf{UTMOS ($\uparrow$)} & \textbf{WER ($\downarrow$)} & \textbf{SBS ($\uparrow$)} & \hspace{1em}\textbf{MOS ($\uparrow$)}\hspace{1em} & \hspace{1em}\textbf{SMOS ($\uparrow$)}\hspace{1em} & \textbf{UTMOS ($\uparrow$)} & \textbf{WER ($\downarrow$)} &\textbf{SBS ($\uparrow$)} \\
\hline
\hline
\textbf{Proposed} & \textbf{4.11$\pm$ \scriptsize{0.09}} & \textbf{4.08$\pm$ \scriptsize{0.10}} & \textbf{3.632} & \textbf{0.211} & \textbf{0.824} & \textbf{4.08$\pm$ \scriptsize{0.13}} & \textbf{3.92$\pm$ \scriptsize{0.11}} & \textbf{3.642} & \textbf{0.241} & \textbf{0.825} \\
\hline

\hspace{1em}-only Wav2Vec 2.0 & 3.79$\pm$ \scriptsize{0.10} & 3.72$\pm$ \scriptsize{0.11} & 2.961 & 0.216 & 0.756 & 3.58$\pm$ \scriptsize{0.11} &3.49$\pm$ \scriptsize{0.11} & 2.872 & 0.274 & 0.738 \\

\hspace{1em}-only BEATs & 3.35$\pm$ \scriptsize{0.08} & 3.53$\pm$ \scriptsize{0.09} & 3.070 & 0.639 & 0.623 & 3.18$\pm$ \scriptsize{0.08} & 3.28$\pm$ \scriptsize{0.10} & 2.909 & 0.688 & 0.606 \\

\hline
\hspace{1em}-w/o parallel & 3.64$\pm$ \scriptsize{0.10} & 3.64$\pm$ \scriptsize{0.09} & 2.898 & 0.505 & 0.706 & 3.34$\pm$ \scriptsize{0.10} & 3.45$\pm$ \scriptsize{0.11} & 3.072 & 0.550 & 0.723 \\
\hline

\hspace{1em}-only AR  & 3.68$\pm$ \scriptsize{0.09} & 3.68$\pm$ \scriptsize{0.09} & 3.200 & 0.295 & 0.811 & 3.62$\pm$ \scriptsize{0.11} & 3.43$\pm$ \scriptsize{0.10} & 3.127 & 0.247 & 0.803 \\
\hline
\end{tabular}
}
\end{table*}

\begin{table*}[!t]
\caption{Ablation Experiments on Chinese Datasets. SBS refers to SpeechBERTScore.}
\label{ablation2}
\centering
\resizebox{1\textwidth}{!}{ 
\begin{tabular}{|l|c|c|c|c|c|c|c|c|c|c|}
\hline
\cline{2-11}
& \multicolumn{5}{c|}{\textbf{Chinese Development Set}} & \multicolumn{5}{c|}{\textbf{Chinese Test Set}} \\
\cline{2-11}
\textbf{Model Name} & \multicolumn{2}{c|}{\textbf{Subjective}} & \multicolumn{3}{c|}{\textbf{Objective}} & \multicolumn{2}{c|}{\textbf{Subjective}} & \multicolumn{3}{c|}{\textbf{Objective}} \\
\cline{2-11}
& \hspace{1em}\textbf{MOS ($\uparrow$)}\hspace{1em} & \hspace{1em}\textbf{SMOS ($\uparrow$)}\hspace{1em} & \textbf{UTMOS ($\uparrow$)} & \textbf{WER ($\downarrow$)} & \textbf{SBS ($\uparrow$)} & \hspace{1em}\textbf{MOS ($\uparrow$)}\hspace{1em} & \hspace{1em}\textbf{SMOS ($\uparrow$)}\hspace{1em} & \textbf{UTMOS ($\uparrow$)} & \textbf{WER ($\downarrow$)} &\textbf{SBS ($\uparrow$)} \\
\hline
\hline
\textbf{Proposed} & \textbf{4.23$\pm$ \scriptsize{0.09}} & \textbf{4.26$\pm$ \scriptsize{0.09}} & {---} & \textbf{0.185} & \textbf{0.799} & \textbf{4.19$\pm$ \scriptsize{0.09}} & \textbf{4.19$\pm$ \scriptsize{0.10}} & {---} & \textbf{0.193} & \textbf{0.825} \\
\hline

\hspace{1em}-only Wav2Vec 2.0 & 3.58$\pm$ \scriptsize{0.12}& 3.48$\pm$ \scriptsize{0.12} & {---}  & 0.235 & 0.648 & 3.48$\pm$ \scriptsize{0.111} & 3.41$\pm$ \scriptsize{0.12}  & {---}  & 0.299 & 0.736 \\

\hspace{1em}-only BEATs & 3.92$\pm$ \scriptsize{0.10} & 3.90$\pm$ \scriptsize{0.11} & {---}  & 0.435 & 0.536 & 3.71$\pm$ \scriptsize{0.11} & 3.75$\pm$ \scriptsize{0.12} & {---}  & 0.627 & 0.556 \\
\hline

\hspace{1em}-w/o parallel & 3.96$\pm$ \scriptsize{0.11} & 3.99$\pm$ \scriptsize{0.10} & {---}  & 0.522 & 0.716 & 4.01$\pm$ \scriptsize{0.11} & 4.03$\pm$ \scriptsize{0.11} & {---}  & 0.298 & 0.652 \\
\hline

\hspace{1em}-only AR  & 3.77$\pm$ \scriptsize{0.10} & 3.89$\pm$ \scriptsize{0.11} & {---}  & 0.299 & 0.699 & 3.92$\pm$ \scriptsize{0.11} & 3.95$\pm$ \scriptsize{0.10} & {---}  & 0.323 & 0.633 \\
\hline
\end{tabular}
}
\end{table*}

\subsection{Ablation Study}
\subsubsection{Ablation Study Settings}
In this section, we conduct detailed ablation experiments. First, we examine parallel tokens combining Wav2Vec 2.0 and BEATs to capture semantics and acoustics simultaneously. We train two tokenizer models using different pre-trained SSL models. The settings \textbf{-only Wav2Vec 2.0} and \textbf{-only BEATs} refer to using Wav2Vec 2.0 and BEATs individually, as opposed to their combined use, to evaluate their performance when used separately. 
Next, we examine the role of parallel models. Parallel AR synthesizes semantics and acoustics simultaneously, with only a temporal linkage. The setting \textbf{-w/o parallel} indicates that semantics and acoustics are merged for RVQ encoding, converting the parallel AR model into a traditional AR framework. Finally, we examine the role of the combined AR+NAR model. The setting \textbf{-only AR} indicates that the AR model is used to predict all RVQ tokens through multiple heads, with six tokens—three for semantic features and three for acoustic features. In this configuration, the speech synthesis model consists solely of the modified Parallel AR model and the Parallel tokenizer decoder.

The experimental setup follows the same procedure as the comparison experiments. Ablation experiments are conducted on both English and Chinese datasets. 

\subsubsection{Ablation Study Analysis}
As shown in the ablation results (Tables \ref{ablation} and \ref{ablation2}), the performance variations across different configurations provide critical insights into the contributions of our proposed components.

First, the comparison between \textbf{-only Wav2Vec 2.0} and \textbf{-only BEATs} reveals a clear trade-off: Wav2Vec 2.0 alone leads to degraded prosody modeling, while BEATs alone suffer from reduced semantic clarity. This observation strongly supports our design choice of a parallel tokenizer, which effectively decouples and jointly optimizes acoustic and semantic representations, thereby mitigating the limitations of single-modality tokenization.

Furthermore, the \textbf{-w/o parallel} setting demonstrates that naively merging semantic and acoustic features for RVQ encoding, similar to traditional TTS approaches, results in inferior prosody preservation. This suggests that entangled representations tend to bias toward semantic information at the expense of fine-grained acoustic details, leading to a drop in SMOS. The decline is particularly pronounced in zero-shot TTS, where disentangled modeling is crucial for capturing speaker-specific characteristics from reference audio.

Finally, \textbf{-only AR} exhibits degraded performance across multiple metrics, including WER and SMOS. This indicates that relying solely on an autoregressive framework to predict all RVQ tokens imposes excessive modeling pressure, especially for long-form speech or highly expressive utterances. The AR model struggles to simultaneously maintain semantic accuracy and acoustic fidelity under constrained capacity, reinforcing the necessity of our hybrid AR+NAR architecture, which strategically distributes the modeling burden for improved robustness.

In summary, our ablation study validates that parallel tokenization, disentangled semantic-acoustic modeling, and hierarchical AR+NAR prediction collectively contribute to the superior performance of Parallel GPT in zero-shot speech synthesis.

\subsection{Validation of Parallel Tokenizer’s Decoupled Performance}

While the disentanglement of semantic and acoustic representations in the parallel tokenizer is critical for effective modeling, no standardized evaluation metric currently exists to assess the separation of semantic and acoustic components quantitatively.

We perform inference-based control experiments to verify the decoupling performance of the Parallel Tokenizer. The trained Parallel Tokenizer considers three branches: semantic, acoustic, and speaker. Removing the semantic input results in unintelligible speech, removing the acoustic input produces bland speech, and removing the speaker vector alters the timbre, making the speech different from the reference. 
Due to the lack of standardized evaluation metrics for speech decoupling experiments, we conduct subjective testing. Evaluators assess the changes in the synthesized speech, and over 90\% agree that the speech exhibits the expected alterations.

\section{Conclusion}

This paper proposes a parallel architecture incorporating autoregressive (AR) and non-autoregressive (NAR) modules. Based on this architecture, we introduce the Parallel Tokenizer and Parallel GPT, a novel model for zero-shot speech synthesis that enables simultaneous processing of acoustic and semantic information, ensuring their independence while improving the synthesized speech’s naturalness, expressiveness, and coherence.
Extensive experiments on both English and Chinese datasets show that Parallel GPT significantly outperforms existing zero-shot TTS models in terms of synthesis quality and efficiency. Our approach addresses key limitations in traditional TTS models and provides a comprehensive framework for generating more dynamic and natural speech.


%

\bibliographystyle{IEEEbib}
\bibliography{refs.bib}

\vfill

\end{document}